\documentclass[twocolumn,showpacs,preprintnumbers,amsmath,amssymb,superscriptaddress]{revtex4}

\usepackage{graphicx}

\newcommand{\ket}[1]{\left|  #1 \right\rangle}

\newcommand{\aver}[1]{\langle {#1} \rangle}
\newcommand{\abs}[1]{\left| {#1} \right|}

\begin{document}

\preprint{XXX}

\title{Interfacing Collective Atomic Excitations and Single Photons}

\author{Jonathan Simon}
\affiliation{Department of Physics, Harvard University, Cambridge,
Massachusetts 02138, USA} \affiliation{Department of Physics,
MIT-Harvard Center for Ultracold Atoms, and Research Laboratory of
Electronics, Massachusetts Institute of Technology, Cambridge,
Massachusetts 02139, USA}

\author{Haruka Tanji}
\affiliation{Department of Physics, Harvard University, Cambridge,
Massachusetts 02138, USA} \affiliation{Department of Physics,
MIT-Harvard Center for Ultracold Atoms, and Research Laboratory of
Electronics, Massachusetts Institute of Technology, Cambridge,
Massachusetts 02139, USA}

\author{James K. Thompson}
\affiliation{Department of Physics, MIT-Harvard Center for Ultracold
Atoms, and Research Laboratory of Electronics, Massachusetts
Institute of Technology, Cambridge, Massachusetts 02139, USA}

\author{Vladan Vuleti\'{c}}
\affiliation{Department of Physics, MIT-Harvard Center for Ultracold
Atoms, and Research Laboratory of Electronics, Massachusetts
Institute of Technology, Cambridge, Massachusetts 02139, USA}

\date{\today}

\begin{abstract}
We study the performance and limitations of a coherent interface
between collective atomic states and single photons. A quantized
spin-wave excitation of an atomic sample inside an optical resonator
is prepared probabilistically, stored, and adiabatically converted
on demand into a sub-Poissonian photonic excitation of the resonator
mode. The measured peak single-quantum conversion efficiency of
$\chi$=0.84(11) and its dependence on various parameters are well
described by a simple model of the mode geometry and multilevel
atomic structure, pointing the way towards implementing
high-performance stationary single-photon sources.
\end{abstract}

\maketitle

A quantum-coherent interface between light and a material structure
that can store quantum states is a pivotal part of a system for
processing quantum information \cite{Zoller05}. In particular, a
quantum memory that can be mapped onto photon number states in a
single spatio-temporal mode could pave the way towards extended
quantum networks \cite{Cirac97,Duan01} and all-optical quantum
computing \cite{Knill01}. While light with sub-Poissonian
fluctuations can be generated by a variety of single-quantum systems
\cite{Brunel99,Lounis00,Darquie05}, a point emitter in free space is
only weakly, and thus irreversibly, coupled to an electromagnetic
continuum.

To achieve reversible coupling, the strength of the emitter-light
interaction can be enhanced by means of an optical resonator, as
demonstrated for quantum dots in the weak-
\cite{Michler00,Santori02}, trapped ions in the intermediate-
\cite{Keller04}, and neutral atoms in the strong-coupling regime
\cite{Kuhn02,McKeever04a}. By controlling the position of a single
atom trapped inside a very-high-finesse resonator, McKeever {\it et
al.} have realized a high-quality deterministic single-photon source
\cite{McKeever04a}. This source operates in principle in the
reversible-coupling regime, although finite mirror losses presently
make it difficult to obtain full reversibility in practice.

Alternatively, superradiant states of an atomic ensemble
\cite{Dicke54} exhibit enhanced coupling to a single electromagnetic
mode. For three-level atoms with two stable ground states these
collective states can be viewed as quantized spin waves, where a
spin-wave quantum (magnon) can be converted into a photon by the
application of a phase-matched laser beam \cite{Duan01}. Such
systems have been used to generate \cite{Chou04,Eisaman04}, store
and retrieve single photons \cite{Chaneliere05,Eisaman05}, to
generate simultaneous-photon pairs \cite{Balic05,Thompson06}, and to
increase the single-photon production rate by feedback
\cite{Matsukevich06,Felinto06,Chen06}. The strong-coupling regime
between magnons and photons can be reached if the sample's optical
depth $OD$ exceeds unity. However, since the failure rate for
magnon-photon conversion in these free-space
\cite{Chou04,Matsukevich04,Eisaman04,Balic05,Chaneliere05,Eisaman05,Laurat06,Matsukevich06,Felinto06,Chen06}
or moderate-finesse-cavity \cite{Black05a,Thompson06} systems has
been around 50\% or higher, which can be realized with $OD \leq 1$,
none of the ensemble systems so far has reached the strong,
reversible-coupling regime.


\begin{figure}
\begin{center}
\includegraphics[width=3.5in,viewport=0 0 700 300]{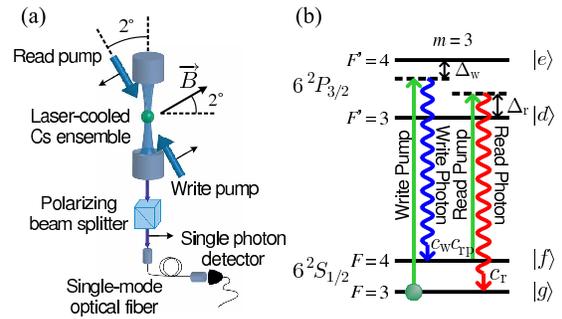}
\end{center}
\caption{(a) Setup for the conditional generation of single photons
using a sample of laser-cooled Cs atoms inside an optical resonator.
(b) Level scheme for the system with hyperfine and magnetic
sublevels $\ket{F,m_F}$. The atomic sample is initially prepared in
$\ket{g}$ by optical pumping.} \label{setup}
\end{figure}

In this Letter, we demonstrate for the first time the
strong-coupling regime between collective spin-wave excitations and
a single electromagnetic mode. This is evidenced by heralded
single-photon generation with a single-quantum conversion efficiency
of $\chi=0.84(11)$, at a seven-fold suppression of two-photon
events. The atomic memory exhibits two Doppler lifetimes, $\tau_s =
230$ ns and $\tau_l$=23 $\mu$s, that are associated with different
magnon wavelengths $\lambda_s$=0.4 $\mu$m and $\lambda_l$=23 $\mu$m
written into the sample.

Our apparatus consists of a 6.6 cm long, standing-wave optical
resonator with a TEM$_{00}$ waist $w_c$=110$\mu$m, finesse
$F$=93(2), linewidth $\kappa/(2\pi)$=24.4(5) MHz , and free spectral
range $\Delta\nu$=2.27 GHz.  The mirror transmissions $M_1,M_2$ and
round-trip loss $L$ near the cesium $D_2$ line wavelength
$\lambda$=$2\pi /k$=852 nm are $M_1$=1.18(2)\%, $M_2$=0.039(2)\%,
and $L$=5.5(1)\%, respectively, such that a photon escapes from the
resonator in the preferred direction with a probability of
$T$=0.175(4). The light exiting from the cavity is
polarization-analyzed, and delivered via a single-mode optical fiber
to a photon counting module. The overall detection probability for a
photon prepared inside the resonator is $q$=$T q_1 q_2
q_3$=2.7(3)\%, which includes photodiode quantum efficiency
$q_1$=0.40(4), interference filter transmission $q_2$=0.609(2), and
fiber coupling and other optical losses $q_3$=0.65(4).  The large
cavity losses in our system arise from Cs deposition.  For the
conditional autocorrelation measurement described at the end of the
paper, we cleaned the mirrors.  This decreased our losses to
$L$=0.30(15)\%, increased the cavity finesse to $F$=420(40) and the
escape probability to $T$=0.78(8), and improved our overall
detection probability to $q$=20(3)\%.

An ensemble containing between $10^3$ and $10^6$ laser-cooled
$^{133}$Cs atoms is prepared along the cavity axis, corresponding to
an adjustable optical depth between $N\eta$=0.1 and $N\eta$=200.
Here $\eta = \eta_0 \abs{c_r}^2$ is the single-atom optical depth
(cooperativity parameter) for the read transition with reduced
dipole matrix element $c_r$=$3/4$ (see Fig. \ref{setup}b), $\eta_0 =
24 F/(\pi k^2 w_c^2)$ is the optical depth for an atom located at a
cavity antinode on a transition with unity matrix element, and $N$
is the effective number of atoms at this location that produces the
same optical depth as the extended sample. The single-atom,
single-photon Rabi frequency $2g$ is given by $\eta = 4g^2/(\kappa
\Gamma)$, where $\Gamma$=$2\pi \times 5.2$ MHz and $\kappa$ are the
atomic and cavity full linewidths, respectively.

Starting with a magneto-optical trap (MOT), we turn off the magnetic
quadrupole field, apply a 1.8 G bias field perpendicular to the
resonator, and optically pump the atoms into a single hyperfine and
magnetic sublevel $\ket{g}$ with two laser beams propagating along
the magnetic field direction. The relevant atomic levels are the
electronic ground states $\ket{g} = \ket{6S_{1/2};F=3,m_F=3}$,
$\ket{f} = \ket{6S_{1/2};4,3}$, and excited states $\ket{e} =
\ket{6P_{3/2};4,3}$, and $\ket{d} = \ket{6P_{3/2};3,3}$ (Fig.
\ref{setup}b). The write and read pump beams, derived from
independent, frequency-stabilized lasers, have a waist size
$w_p$=300 $\mu$m, enclose a small angle $\theta \approx 2^\circ$
with the cavity axis, and are linearly polarized along the bias
field (Fig. \ref{setup}a). The write pump is applied for 60 ns with
a detuning of $\Delta_w/(2\pi)=-40$MHz from the $\ket{g} \rightarrow
\ket{e}$ transition at a typical intensity of 70 mW/cm$^2$. With
some small probability a ``write'' photon is generated inside the
resonator by spontaneous Raman scattering on the $\ket{g}
\rightarrow \ket{e} \rightarrow \ket{f}$ transition to which a
resonator TEM$_{00}$ mode is tuned \cite{Duan01,Black05a}. At some
later time, the quantized spin wave generated in the write process
is strongly (superradiantly) coupled to the cavity if the Raman
emission $\ket{f} \rightarrow \ket{d} \rightarrow \ket{g}$ from a
phase-matched read pump beam restores the sample's initial momentum
distribution \cite{Duan01,Dicke54,Black05a}. The read pump is ramped
on in 100 ns, with a peak intensity of up to 7 W/cm$^2$. It is
detuned by $\Delta_r/(2 \pi)$=60 MHz relative to the $\ket{f}
\rightarrow \ket{d}$ transition, such that the ``read'' photon is
emitted into another TEM$_{00}$ resonator mode. The write-read
process is repeated for 2 ms (up to 800 times) per MOT cycle of 100
ms.

As the conversion efficiency $\chi$ of a single stored magnon into a
photon in the cavity approaches unity, small fractional
uncertainties in $\chi$ result in large uncertainties in the failure
rate $1-\chi$.  However, the interesting physics of entangled atomic
states coupling to photons that rules the matter-light interface
hinges on understanding the failure rate.  Thus, we explore how to
accurately estimate $\chi$ by studying the directly measurable
conditional retrieval efficiency $R_c=(\aver{n_w n_r} - \aver{n_w}
\aver{n_r}) / \aver{n_w}$, and unconditional retrieval efficiency
$R_u=\aver{n_r} / \aver{n_w}$.  Here $n_w$ and $n_r$ are the write
and read photon numbers in a given time interval, respectively,
referenced to within the resonator. Note that neither measure $R_c,
R_u$ is a priori an accurate estimator of the single-quantum
conversion efficiency $\chi$.  The conditional quantity $R_c$ is
insensitive to read backgrounds, but requires accurate calibration
of detection efficiency, and systematically differs from $\chi$ both
at low and high $\aver{n_w}$ \cite{Laurat06}. $R_u$ can be measured
at larger $\aver{n_w}$ and provides better statistics since it does
not rely on correlated events, but is sensitive to read backgrounds
which must be independently measured, e.g., by breaking the
phase-matching condition \cite{Black05a}.

\begin{figure}
\begin{center}
\includegraphics[width=3.5in]{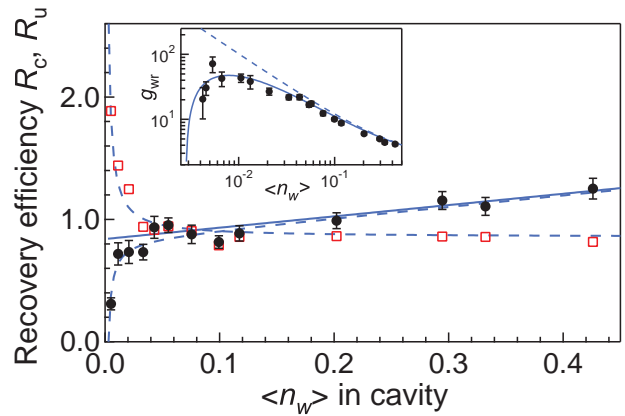}
\end{center}
\caption{Conditional ($R_c$, solid circles) and unconditional
($R_u$, open squares) retrieval with model predictions, versus
intracavity write photon number $\aver{n_w}$, at a write-read delay
of 80ns. The single-quantum conversion efficiency $\chi$ can also be
obtained as the $y$-axis intercept of the linear fit to $R_c$ (solid
black line). Inset: Non-classical write-read correlation $g_{wr}>2$
with model (solid line) and theoretical limit $g_{wr} \leq
1/\aver{n_w}$ (dashed line).} \label{Rvsnw}
\end{figure}

Fig. \ref{Rvsnw} shows the conditional and unconditional retrieval
efficiencies $R_c, R_u$ versus average write photon number
$\aver{n_w}$ inside the lower-finesse resonator at fixed optical
depth $N\eta$=10.
A carefully calibrated $17(4)\%$ correction due to detector
afterpulsing has been applied to $R_c$. The rise in $R_u$ at small
$\aver{n_w}$ is due to read backgrounds, while the drop in $R_c$ is
due to write backgrounds, that are not accompanied by a spin wave.
The increase of $R_c$ with $\aver{n_w}$ is due to double
excitations.
An accurate value for the single-quantum conversion efficiency
$\chi$ can be extracted from the measured data by means of a model
that includes uncorrelated constant write and read backgrounds,
independently measured to be $b_w = 0.0028(4)$ and $b_r = 0.0074(9)$
when referenced to the cavity.
This model predicts $\aver{n_w} R_u =\chi (\aver{n_w}-b_w)+ b_r$ and
$\aver{n_w} R_c = \chi (\aver{n_w}-b_w) [1+(g_{ww}-1)
(\aver{n_w}-b_w)]$. The measured write second-order autocorrelation
function $g_{ww}^{meas}=2.4(2)$ differs from the expected value
$g_{ww}=2$, likely due to observed fluctuations in write pulse
intensity. A fit of $R_c, R_u$ to the model, with the conversion
$\chi$ as the only fitting parameter, yields a good match between
data and model, and good agreement between the value $\chi_c =
0.84(11)$ extracted from the conditional and the value $\chi_u =
0.85(2)$ extracted from the unconditional retrieval efficiency.
$\chi_u$, being independent of detection efficiency, is more
precise.
Since $b_w, b_r \ll 1$, the magnon-photon conversion $\chi$ can also
be estimated as the $y$ intercept of the linear fit $R_c =
\chi(1+(g_{ww}-1)\aver{n_w})$. The inset to Fig. \ref{Rvsnw} shows
the write-read cross correlation $g_{wr} = \aver{n_w n_r}/
(\aver{n_w} \aver{n_r})$ versus $\aver{n_w}$, as well as the
predicted dependence with no free parameters. In the region
$\aver{n_w}>0.05$, where $g_{wr}$ approaches its fundamental limit
$g_{wr} \leq 1/ \aver{n_w}$, backgrounds are negligible, and the
unconditional recovery $R_u$ is also a good estimate of $\chi$. In
the figures which follow, we estimate $\chi$ as
$R_c/(1+\aver{n_w})$, and use it to examine the physical limitations
on the magnon-photon interface.

The most fundamental limit on the conversion process
$\chi_0=N\eta/(N\eta +1)$ arises from the competition between the
sample's collective coupling to the cavity mode, and single-atom
emission into free space.  In the off-resonant
(collective-scattering) regime this limit originates from the
collective enhancement of the read rate by a factor $N \eta$
relative to the single-atom free-space scattering rate
\cite{Black05a}. In the on-resonance (dark-state rotation) regime
\cite{Duan01,Kuhn02,McKeever04a} the limit $\chi_0$ is due to the
stronger suppression of free-space scattering (by a factor
$(N\eta)^{-2}$) compared to the suppression of cavity emission
(factor $(N\eta)^{-1}$). In either case, large optical depth is key
to a good interface.

\begin{figure}
\begin{center}
\includegraphics[width=3.0in]{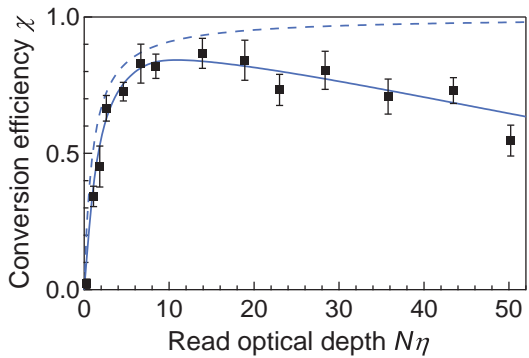}
\end{center}
\caption{Magnon-photon conversion efficiency $\chi$ versus read
optical depth $N\eta$, at a write-read delay of 120 ns.  The optical
depth is extracted from the write scattering rate and known
intensities and detunings.  The dashed line shows the predicted
conversion $\chi_0$ for a three-level system, the solid line is the
prediction from a model including dephasing from additional excited
states.} \label{RvsNeta}
\end{figure}

To the extent that the atomic system is not a simple three-level
system,  additional sources of magnon decoherence, such as
non-resonant scattering from other excited states, reduce the
conversion efficiency. More relevant in the present case are
spatially inhomogeneous light shifts due to other excited states
that decrease linearly, rather than quadratically, with the
excited-state energy splittings. Such light shifts dephase the spin
grating, and reduce the magnon-photon conversion by $\chi_{ls} =
(1-2 s^4 \phi_r^2)$ to lowest order in the ratio $s = w_c/w_p \ll
1$. Here $\phi_r$ is the average light-shift-induced phase
accumulated by an atom on the pump beam axis during the read
process, and $w_c$ ($w_p$) is the cavity (read pump) waist. Note
that $\chi_{ls}$ depends upon read pump size $w_p$, but not the read
pump intensity $I_r$, since both light shift and read rate are
proportional to $I_r$.  As such, as the pump waist is increased at
fixed read intensity, the resulting failure rate should decrease as
the inverse square of the pump power.

Fig. \ref{RvsNeta} shows that this dephasing effect dramatically
changes the dependence of the conversion efficiency on optical depth
$N\eta$. While the conversion efficiency $\chi_0$ for a
three-level-atom approaches unity for large optical depth $N\eta$
(dashed line), the increase in read-photon emission time in the
dark-state rotation regime (by a factor $N\eta$) for atoms with
multiple excited states increases the dephasing $\chi_{ls}$, and
reduces the conversion. The predicted conversion $\chi_0 \chi_{ls}$
including all atomic excited hyperfine states produces the correct
functional form, as well as the position and peak value of the
recovery efficiency, at a waist ratio of $s^{-1}$=$w_p/w_c$=3, in
good agreement with the measured value of 3.0(4).

\begin{figure}
\begin{center}
\includegraphics[width=3.0in]{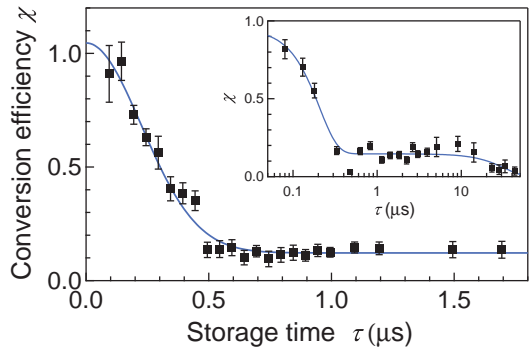}
\end{center}
\caption{Conditional single-photon conversion efficiency $\chi$
versus the delay time between write and read pulses $\tau$.  The two
time scales, as apparent in the inset, are due to the superposition
of a short- and a long-wavelength magnon in the standing-wave
resonator.} \label{RvsDelayTotal}
\end{figure}

The prediction in Fig. \ref{RvsNeta} also includes a small
conversion reduction due to the decoherence of the magnon caused by
the atoms' thermal motion during the 120 ns storage time. For the
small angle $\theta \approx 2^\circ$ between running-wave pump beams
and cavity standing wave, the write photon emission process creates
a superposition of two spin waves of very different wavelengths.
Backward emission corresponds to a short wavelength $\lambda_s
\approx \lambda/2 =$ 0.4 $\mu$m, and is highly Doppler-sensitive,
while forward emission with $\lambda_l = \lambda/(2 \sin(\theta/2))$
= 23 $\mu$m is nearly Doppler free. The recovery versus storage time
$\tau$ at $N\eta = 10$ (Fig. \ref{RvsDelayTotal}) shows the two
corresponding Gaussian time constants $\tau_s=240$ ns and
$\tau_l=23$ $\mu$s.

The long-time conversion is limited to 25\%, because each individual
spin-wave component alone can only be recovered with 50\%
probability due to the mismatch between the standing wave cavity
mode and the running-wave magnon. The highest observed conversion
efficiency in Fig. \ref{RvsDelayTotal} of $\chi $=0.95(13) is higher
than for the inset or Fig. \ref{Rvsnw}.  The data for Fig.
\ref{RvsDelayTotal} was taken after carefully realigning the bias
magnetic field along the quantization axis defined by the pump beam
polarizations, while the inset and Fig. \ref{Rvsnw} were taken
before realignment. This suggests that spin precession due to
imperfect magnetic field alignment could also reduce the conversion
efficiency.  The result $\chi $=0.95 was obtained for a single
write-photon value $\aver{n_w} $=0.27(3), so we conservatively quote
$\chi $=0.84 obtained from the fit to the data versus $n_w$ with
un-optimized fields in Fig. \ref{Rvsnw}.

Using the lower loss ($F$=420) cavity with clean mirrors to minimize
sensitivity to detector afterpulsing and improve the data collection
rate, we measure the read autocorrelation function $g_{rr|w}$
conditioned on having detected a write photon. Due to the seven
times higher detection efficiency, the detector dark count rate does
not appreciably lower the recovery down to $n_w=0.005$ in this
configuration. For $\aver{n_w}$=0.007 we obtain $g_{rr|w}$=0.15(8)
at an optical depth $N\eta$=10, clearly demonstrating the
sub-Poissonian nature of the source.

In summary, we have realized an interface between spin-wave quanta
and narrowband single photons with a performance near 90\%,
representing the first experimental demonstration of strong coupling
between collective spin-wave excitations and photons. Several
proposed mechanisms appear to adequately explain the remaining
failure rate of the magnon-photon interface, and indicate the path
to future improvements.
If the conditional single-photon source described here were operated
as a single-trial source by applying the read beam only when a write
photon was detected, it already would have almost comparable
performance to recently demonstrated feedback-enhanced sources
\cite{Matsukevich06,Felinto06,Chen06}: at $n_w =0.007$ our source
would unconditionally deliver photons with probability 0.6\% at
$g_{rr}=0.15$, to be compared to 5.4\% at $g_{rr}=0.41$ for 150
trials \cite{Matsukevich06}, or 2.5\% at $g_{rr}=0.3$ \cite{Chen06}
for 12 trials. It should be straightforward to implement feedback as
demonstrated in Refs. \cite{Matsukevich06,Felinto06,Chen06} into our
setup. If the Doppler effect can be eliminated by confining the
atoms in a far-detuned optical lattice, the resulting substantial
increase in magnon storage time would allow the 150 trials necessary
to implement an unconditional source with near-unity single-photon
probability.

This work was supported in parts by the NSF, DARPA and ARO. J.S.
acknowledges NDSEG and NSF fellowships.

\begin{thebibliography}{25}
\expandafter\ifx\csname
natexlab\endcsname\relax\def\natexlab#1{#1}\fi
\expandafter\ifx\csname bibnamefont\endcsname\relax
  \def\bibnamefont#1{#1}\fi
\expandafter\ifx\csname bibfnamefont\endcsname\relax
  \def\bibfnamefont#1{#1}\fi
\expandafter\ifx\csname citenamefont\endcsname\relax
  \def\citenamefont#1{#1}\fi
\expandafter\ifx\csname url\endcsname\relax
  \def\url#1{\texttt{#1}}\fi
\expandafter\ifx\csname urlprefix\endcsname\relax\def\urlprefix{URL
}\fi \providecommand{\bibinfo}[2]{#2}
\providecommand{\eprint}[2][]{\url{#2}}

\bibitem[{\citenamefont{Zoller}(2005)}]{Zoller05}
\bibinfo{author}{\bibfnamefont{P.}~\bibnamefont{Zoller} {\it et al.}},
  \bibinfo{journal}{Euro.\ Phys.\ J.\ D} \textbf{\bibinfo{volume}{36}},
  \bibinfo{pages}{203} (\bibinfo{year}{2005}).

\bibitem[{\citenamefont{Cirac et~al.}(1997)\citenamefont{Cirac, Zoller, Kimble,
  and Mabuchi}}]{Cirac97}
\bibinfo{author}{\bibfnamefont{J.~I.} \bibnamefont{Cirac}},
  \bibinfo{author}{\bibfnamefont{P.}~\bibnamefont{Zoller}},
  \bibinfo{author}{\bibfnamefont{H.~J.} \bibnamefont{Kimble}},
  \bibnamefont{and} \bibinfo{author}{\bibfnamefont{H.}~\bibnamefont{Mabuchi}},
  \bibinfo{journal}{Phys.\ Rev.\ Lett.} \textbf{\bibinfo{volume}{78}},
  \bibinfo{pages}{3221} (\bibinfo{year}{1997}).

\bibitem[{\citenamefont{Duan et~al.}(2001)\citenamefont{Duan, Lukin, Cirac, and
  Zoller}}]{Duan01}
\bibinfo{author}{\bibfnamefont{L.-M.} \bibnamefont{Duan}},
  \bibinfo{author}{\bibfnamefont{M.~D.} \bibnamefont{Lukin}},
  \bibinfo{author}{\bibfnamefont{J.~I.} \bibnamefont{Cirac}}, \bibnamefont{and}
  \bibinfo{author}{\bibfnamefont{P.}~\bibnamefont{Zoller}},
  \bibinfo{journal}{Nature} \textbf{\bibinfo{volume}{414}},
  \bibinfo{pages}{413} (\bibinfo{year}{2001}).

\bibitem[{\citenamefont{Knill et~al.}(2001)\citenamefont{Knill, Laflamme, and
  Milburn}}]{Knill01}
\bibinfo{author}{\bibfnamefont{E.}~\bibnamefont{Knill}},
  \bibinfo{author}{\bibfnamefont{R.}~\bibnamefont{Laflamme}}, \bibnamefont{and}
  \bibinfo{author}{\bibfnamefont{G.}~\bibnamefont{Milburn}},
  \bibinfo{journal}{Nature} \textbf{\bibinfo{volume}{409}}, \bibinfo{pages}{46}
  (\bibinfo{year}{2001}).

\bibitem[{\citenamefont{Brunel et~al.}(1999)\citenamefont{Brunel, Lounis,
  Tamarat, and Orrit}}]{Brunel99}
\bibinfo{author}{\bibfnamefont{C.}~\bibnamefont{Brunel}},
  \bibinfo{author}{\bibfnamefont{B.}~\bibnamefont{Lounis}},
  \bibinfo{author}{\bibfnamefont{P.}~\bibnamefont{Tamarat}}, \bibnamefont{and}
  \bibinfo{author}{\bibfnamefont{M.}~\bibnamefont{Orrit}},
  \bibinfo{journal}{Phys.\ Rev.\ Lett.} \textbf{\bibinfo{volume}{83}},
  \bibinfo{pages}{2722} (\bibinfo{year}{1999}).

\bibitem[{\citenamefont{Lounis and Moerner}(2000)}]{Lounis00}
\bibinfo{author}{\bibfnamefont{B.}~\bibnamefont{Lounis}} \bibnamefont{and}
  \bibinfo{author}{\bibfnamefont{W.~E.} \bibnamefont{Moerner}},
  \bibinfo{journal}{Nature} \textbf{\bibinfo{volume}{407}},
  \bibinfo{pages}{491} (\bibinfo{year}{2000}).

\bibitem[{\citenamefont{Darqui\'{e} et~al.}(2005)\citenamefont{Darqui\'{e},
  Jones, Dingjan, Beugnon, Bergamini, Sortais, Messin, Browaeys, and
  Grangier}}]{Darquie05}
\bibinfo{author}{\bibfnamefont{B.}~\bibnamefont{Darqui\'{e}}},
  \bibinfo{author}{\bibfnamefont{M.~P.~A.} \bibnamefont{Jones}},
  \bibinfo{author}{\bibfnamefont{J.}~\bibnamefont{Dingjan}},
  \bibinfo{author}{\bibfnamefont{J.}~\bibnamefont{Beugnon}},
  \bibinfo{author}{\bibfnamefont{S.}~\bibnamefont{Bergamini}},
  \bibinfo{author}{\bibfnamefont{Y.}~\bibnamefont{Sortais}},
  \bibinfo{author}{\bibfnamefont{G.}~\bibnamefont{Messin}},
  \bibinfo{author}{\bibfnamefont{A.}~\bibnamefont{Browaeys}}, \bibnamefont{and}
  \bibinfo{author}{\bibfnamefont{P.}~\bibnamefont{Grangier}},
  \bibinfo{journal}{Science} \textbf{\bibinfo{volume}{309}},
  \bibinfo{pages}{454} (\bibinfo{year}{2005}).

\bibitem[{\citenamefont{Michler et~al.}(2000)\citenamefont{Michler, Kiraz,
  Becher, Schoenfeld, Petroff, Zhang, Hu, and Imamoglu}}]{Michler00}
\bibinfo{author}{\bibfnamefont{P.}~\bibnamefont{Michler}},
  \bibinfo{author}{\bibfnamefont{A.}~\bibnamefont{Kiraz}},
  \bibinfo{author}{\bibfnamefont{C.}~\bibnamefont{Becher}},
  \bibinfo{author}{\bibfnamefont{W.~V.} \bibnamefont{Schoenfeld}},
  \bibinfo{author}{\bibfnamefont{P.~M.} \bibnamefont{Petroff}},
  \bibinfo{author}{\bibfnamefont{L.}~\bibnamefont{Zhang}},
  \bibinfo{author}{\bibfnamefont{E.}~\bibnamefont{Hu}}, \bibnamefont{and}
  \bibinfo{author}{\bibfnamefont{A.}~\bibnamefont{Imamoglu}},
  \bibinfo{journal}{Science} \textbf{\bibinfo{volume}{290}},
  \bibinfo{pages}{2282} (\bibinfo{year}{2000}).

\bibitem[{\citenamefont{Santori et~al.}(2002)\citenamefont{Santori, Fattal,
  Vuckovic, Solomon, and Yamamoto}}]{Santori02}
\bibinfo{author}{\bibfnamefont{C.}~\bibnamefont{Santori}},
  \bibinfo{author}{\bibfnamefont{D.}~\bibnamefont{Fattal}},
  \bibinfo{author}{\bibfnamefont{J.}~\bibnamefont{Vuckovic}},
  \bibinfo{author}{\bibfnamefont{G.~S.} \bibnamefont{Solomon}},
  \bibnamefont{and} \bibinfo{author}{\bibfnamefont{Y.}~\bibnamefont{Yamamoto}},
  \bibinfo{journal}{Nature} \textbf{\bibinfo{volume}{419}},
  \bibinfo{pages}{594} (\bibinfo{year}{2002}).

\bibitem[{\citenamefont{Keller et~al.}(2004)\citenamefont{Keller, Lange,
  Hayasaka, Lange, and Walther}}]{Keller04}
\bibinfo{author}{\bibfnamefont{M.}~\bibnamefont{Keller}},
  \bibinfo{author}{\bibfnamefont{B.}~\bibnamefont{Lange}},
  \bibinfo{author}{\bibfnamefont{K.}~\bibnamefont{Hayasaka}},
  \bibinfo{author}{\bibfnamefont{W.}~\bibnamefont{Lange}}, \bibnamefont{and}
  \bibinfo{author}{\bibfnamefont{H.}~\bibnamefont{Walther}},
  \bibinfo{journal}{Nature} \textbf{\bibinfo{volume}{431}},
  \bibinfo{pages}{1075} (\bibinfo{year}{2004}).

\bibitem[{\citenamefont{Kuhn et~al.}(2002)\citenamefont{Kuhn, Hennrich, and
  Rempe}}]{Kuhn02}
\bibinfo{author}{\bibfnamefont{A.}~\bibnamefont{Kuhn}},
  \bibinfo{author}{\bibfnamefont{M.}~\bibnamefont{Hennrich}}, \bibnamefont{and}
  \bibinfo{author}{\bibfnamefont{G.}~\bibnamefont{Rempe}},
  \bibinfo{journal}{Phys.\ Rev.\ Lett.} \textbf{\bibinfo{volume}{89}},
  \bibinfo{pages}{067901} (\bibinfo{year}{2002}).

\bibitem[{\citenamefont{McKeever et~al.}(2004)\citenamefont{McKeever, Boca,
  Boozer, Miller, Buck, Kuzmich, and Kimble}}]{McKeever04a}
\bibinfo{author}{\bibfnamefont{J.}~\bibnamefont{McKeever}},
  \bibinfo{author}{\bibfnamefont{A.}~\bibnamefont{Boca}},
  \bibinfo{author}{\bibfnamefont{A.}~\bibnamefont{Boozer}},
  \bibinfo{author}{\bibfnamefont{R.}~\bibnamefont{Miller}},
  \bibinfo{author}{\bibfnamefont{J.}~\bibnamefont{Buck}},
  \bibinfo{author}{\bibfnamefont{A.}~\bibnamefont{Kuzmich}}, \bibnamefont{and}
  \bibinfo{author}{\bibfnamefont{H.}~\bibnamefont{Kimble}},
  \bibinfo{journal}{Science} \textbf{\bibinfo{volume}{303}},
  \bibinfo{pages}{1992} (\bibinfo{year}{2004}).

\bibitem[{\citenamefont{Dicke}(1954)}]{Dicke54}
\bibinfo{author}{\bibfnamefont{R.~H.} \bibnamefont{Dicke}},
  \bibinfo{journal}{Phys.\ Rev.} \textbf{\bibinfo{volume}{93}},
  \bibinfo{pages}{99} (\bibinfo{year}{1954}).

\bibitem[{\citenamefont{Chou et~al.}(2004)\citenamefont{Chou, Polyakov,
  Kuzmich, and Kimble}}]{Chou04}
\bibinfo{author}{\bibfnamefont{C.~W.} \bibnamefont{Chou}},
  \bibinfo{author}{\bibfnamefont{S.~V.} \bibnamefont{Polyakov}},
  \bibinfo{author}{\bibfnamefont{A.}~\bibnamefont{Kuzmich}}, \bibnamefont{and}
  \bibinfo{author}{\bibfnamefont{H.~J.} \bibnamefont{Kimble}},
  \bibinfo{journal}{Phys.\ Rev.\ Lett.} \textbf{\bibinfo{volume}{92}},
  \bibinfo{pages}{213601} (\bibinfo{year}{2004}).

\bibitem[{\citenamefont{Matsukevich and Kuzmich}(2004)}]{Matsukevich04}
\bibinfo{author}{\bibfnamefont{D.}~\bibnamefont{Matsukevich}} \bibnamefont{and}
  \bibinfo{author}{\bibfnamefont{A.}~\bibnamefont{Kuzmich}},
  \bibinfo{journal}{Science} \textbf{\bibinfo{volume}{306}},
  \bibinfo{pages}{663} (\bibinfo{year}{2004}).

\bibitem[{\citenamefont{Eisaman et~al.}(2004)\citenamefont{Eisaman, Childress,
  Andre, Massou, Zibrov, and Lukin}}]{Eisaman04}
\bibinfo{author}{\bibfnamefont{M.~D.} \bibnamefont{Eisaman}},
  \bibinfo{author}{\bibfnamefont{L.}~\bibnamefont{Childress}},
  \bibinfo{author}{\bibfnamefont{A.}~\bibnamefont{Andre}},
  \bibinfo{author}{\bibfnamefont{F.}~\bibnamefont{Massou}},
  \bibinfo{author}{\bibfnamefont{A.~S.} \bibnamefont{Zibrov}},
  \bibnamefont{and} \bibinfo{author}{\bibfnamefont{M.~D.} \bibnamefont{Lukin}},
  \bibinfo{journal}{Phys.\ Rev.\ Lett.} \textbf{\bibinfo{volume}{93}},
  \bibinfo{pages}{233602} (\bibinfo{year}{2004}).

\bibitem[{\citenamefont{Bali\'{c} et~al.}(2005)\citenamefont{Bali\'{c}, Braje,
  Kolchin, Yin, and Harris}}]{Balic05}
\bibinfo{author}{\bibfnamefont{V.}~\bibnamefont{Bali\'{c}}},
  \bibinfo{author}{\bibfnamefont{D.A.}~\bibnamefont{Braje}},
  \bibinfo{author}{\bibfnamefont{P.}~\bibnamefont{Kolchin}},
  \bibinfo{author}{\bibfnamefont{G.Y.}~\bibnamefont{Yin}}, \bibnamefont{and}
  \bibinfo{author}{\bibfnamefont{S.E.}~\bibnamefont{Harris}},
  \bibinfo{journal}{Phys.\ Rev.\ Lett.} \textbf{\bibinfo{volume}{94}},
  \bibinfo{pages}{183601} (\bibinfo{year}{2005}).

\bibitem[{\citenamefont{Chaneliere et~al.}(2005)\citenamefont{Chaneliere,
  Matsukevich, Jenkins, Lan, Kennedy, and Kuzmich}}]{Chaneliere05}
\bibinfo{author}{\bibfnamefont{T.}~\bibnamefont{Chaneliere}},
  \bibinfo{author}{\bibfnamefont{D.~N.} \bibnamefont{Matsukevich}},
  \bibinfo{author}{\bibfnamefont{S.~D.} \bibnamefont{Jenkins}},
  \bibinfo{author}{\bibfnamefont{S.~Y.} \bibnamefont{Lan}},
  \bibinfo{author}{\bibfnamefont{T.~A.~B.} \bibnamefont{Kennedy}},
  \bibnamefont{and} \bibinfo{author}{\bibfnamefont{A.}~\bibnamefont{Kuzmich}},
  \bibinfo{journal}{Nature} \textbf{\bibinfo{volume}{438}},
  \bibinfo{pages}{833} (\bibinfo{year}{2005}).

\bibitem[{\citenamefont{Eisaman et~al.}(2005)\citenamefont{Eisaman, Andr\'{e},
  Massou, Fleischhauer, Zibrov, and Lukin}}]{Eisaman05}
\bibinfo{author}{\bibfnamefont{M.~D.} \bibnamefont{Eisaman}},
  \bibinfo{author}{\bibfnamefont{A.}~\bibnamefont{Andr\'{e}}},
  \bibinfo{author}{\bibfnamefont{F.}~\bibnamefont{Massou}},
  \bibinfo{author}{\bibfnamefont{M.}~\bibnamefont{Fleischhauer}},
  \bibinfo{author}{\bibfnamefont{A.~S.} \bibnamefont{Zibrov}},
  \bibnamefont{and} \bibinfo{author}{\bibfnamefont{M.~D.} \bibnamefont{Lukin}},
  \bibinfo{journal}{Nature} \textbf{\bibinfo{volume}{438}},
  \bibinfo{pages}{837} (\bibinfo{year}{2005}).

\bibitem[{\citenamefont{Laurat et~al.}(2006)\citenamefont{Laurat, Riedmatten,
  Felinto, Chou, Schomburg, and Kimble}}]{Laurat06}
\bibinfo{author}{\bibfnamefont{J.}~\bibnamefont{Laurat}},
  \bibinfo{author}{\bibfnamefont{H.}~\bibnamefont{Riedmatten}},
  \bibinfo{author}{\bibfnamefont{D.}~\bibnamefont{Felinto}},
  \bibinfo{author}{\bibfnamefont{C.~W.} \bibnamefont{Chou}},
  \bibinfo{author}{\bibfnamefont{E.~W.} \bibnamefont{Schomburg}},
  \bibnamefont{and} \bibinfo{author}{\bibfnamefont{H.~J.}
  \bibnamefont{Kimble}}, \bibinfo{journal}{Optics Express}
  \textbf{\bibinfo{volume}{14}}, \bibinfo{pages}{6912} (\bibinfo{year}{2006}).

\bibitem[{\citenamefont{Matsukevich et~al.}(2006)\citenamefont{Matsukevich,
  Chaneliere, Jenkins, Lan, Kennedy, and Kuzmich}}]{Matsukevich06}
\bibinfo{author}{\bibfnamefont{D.~N.} \bibnamefont{Matsukevich}},
  \bibinfo{author}{\bibfnamefont{T.}~\bibnamefont{Chaneliere}},
  \bibinfo{author}{\bibfnamefont{S.~D.} \bibnamefont{Jenkins}},
  \bibinfo{author}{\bibfnamefont{S.~Y.} \bibnamefont{Lan}},
  \bibinfo{author}{\bibfnamefont{T.~A.~B.} \bibnamefont{Kennedy}},
  \bibnamefont{and} \bibinfo{author}{\bibfnamefont{A.}~\bibnamefont{Kuzmich}},
  \bibinfo{journal}{Phys.\ Rev.\ Lett.} \textbf{\bibinfo{volume}{97}},
  \bibinfo{pages}{013601} (\bibinfo{year}{2006}).

\bibitem[{\citenamefont{Felinto et~al.}(2006)\citenamefont{Felinto, Chou, Laurat, Schomburg, de~Riedmatten, and Kimble}}]{Felinto06}
\bibinfo{author}{\bibinfo{author}{\bibfnamefont{D.}~\bibnamefont{Felinto}},
  \bibinfo{author}{\bibfnamefont{C.~W.} \bibnamefont{Chou}},
  \bibinfo{author}{\bibfnamefont{J.}~\bibnamefont{Laurat}},
  \bibinfo{author}{\bibfnamefont{E.~W.} \bibnamefont{Schomburg}},
  \bibfnamefont{H.}~\bibnamefont{de~Riedmatten}},
   \bibnamefont{and}
  \bibinfo{author}{\bibfnamefont{H.~J.} \bibnamefont{Kimble}},
  \bibinfo{journal}{Nature Physics} \textbf{\bibinfo{volume}{2}},
  \bibinfo{pages}{844} (\bibinfo{year}{2006}).

\bibitem[{\citenamefont{Chen et~al.}(2006)\citenamefont{Chen, Chen, Strassel,
  Yuan, Zhao, Schmiedmayer, and Pan}}]{Chen06}
\bibinfo{author}{\bibfnamefont{S.}~\bibnamefont{Chen}},
  \bibinfo{author}{\bibfnamefont{Y.-A.} \bibnamefont{Chen}},
  \bibinfo{author}{\bibfnamefont{T.}~\bibnamefont{Strassel}},
  \bibinfo{author}{\bibfnamefont{Z.-S.} \bibnamefont{Yuan}},
  \bibinfo{author}{\bibfnamefont{B.}~\bibnamefont{Zhao}},
  \bibinfo{author}{\bibfnamefont{J.}~\bibnamefont{Schmiedmayer}},
  \bibnamefont{and} \bibinfo{author}{\bibfnamefont{J.-W.} \bibnamefont{Pan}},
  \bibinfo{journal}{Phys.\ Rev.\ Lett.} \textbf{\bibinfo{volume}{97}},
  \bibinfo{pages}{173004} (\bibinfo{year}{2006}).

\bibitem[{\citenamefont{Black et~al.}(2005)\citenamefont{Black, Thompson, and
  Vuleti\'{c}}}]{Black05a}
\bibinfo{author}{\bibfnamefont{A.~T.} \bibnamefont{Black}},
  \bibinfo{author}{\bibfnamefont{J.~K.} \bibnamefont{Thompson}},
  \bibnamefont{and}
  \bibinfo{author}{\bibfnamefont{V.}~\bibnamefont{Vuleti\'{c}}},
  \bibinfo{journal}{Phys.\ Rev.\ Lett.} \textbf{\bibinfo{volume}{95}},
  \bibinfo{pages}{133601} (\bibinfo{year}{2005}).

\bibitem[{\citenamefont{Thompson et~al.}(2006)\citenamefont{Thompson, Simon,
  Loh, and Vuletic}}]{Thompson06}
\bibinfo{author}{\bibfnamefont{J.~K.} \bibnamefont{Thompson}},
  \bibinfo{author}{\bibfnamefont{J.}~\bibnamefont{Simon}},
  \bibinfo{author}{\bibfnamefont{H.-Q.} \bibnamefont{Loh}}, \bibnamefont{and}
  \bibinfo{author}{\bibfnamefont{V.}~\bibnamefont{Vuletic}},
  \bibinfo{journal}{Science} \textbf{\bibinfo{volume}{313}},
  \bibinfo{pages}{74} (\bibinfo{year}{2006}).

\end{thebibliography}

\end{document}